\documentclass[journal=jacsat,manuscript=article]{achemso}

\usepackage[version=3]{mhchem} 
\usepackage{xcolor}
\usepackage{soul}

\author{Pierre-Luc Thériault}
    \affiliation{Polytechnique Montréal, Department of Engineering Physics, Montréal H3T 1J4, QC, Canada}
    \altaffiliation{These authors contributed equally}
\author{Heorhii V. Humeniuk}
    \affiliation{McGill University, Department of Chemistry, Montréal H3A 0B8, QC, Canada}
    \altaffiliation{These authors contributed equally}
\author{Zhechang He}
    \affiliation{McGill University, Department of Chemistry, Montréal H3A 0B8, QC, Canada}
\author{Gabriel Juteau}
    \affiliation{Polytechnique Montréal, Department of Engineering Physics, Montréal H3T 1J4, QC, Canada}  
\author{Alexandre Malinge}
    \affiliation{Polytechnique Montréal, Department of Engineering Physics, Montréal H3T 1J4, QC, Canada}  
\author{Dmytro F. Perepichka}
    \email{dmytro.perepichka@mcgill.ca}
    \affiliation{McGill University, Department of Chemistry, Montréal H3A 0B8, QC, Canada}
\author{Stéphane Kéna-Cohen}
    \email{s.kena-cohen@polymtl.ca}
    \affiliation{Polytechnique Montréal, Department of Engineering Physics, Montréal H3T 1J4, QC, Canada}

\title[An \textsf{achemso} demo]
  {Molecular Engineering for Enhanced Second-Order Nonlinear Response in Spontaneously-Oriented Evaporated Organic Films}


\begin{document}

\begin{tocentry}

\includegraphics[width=50mm]{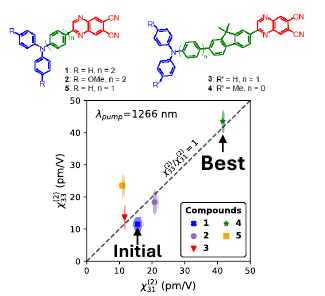}
Intramolecular bridge-locking frustrates detrimental anti-parallel packing in evaporated organic films. This design enhances net polar orientation,leading to a twofold off-resonance increase in the second-order nonlinear susceptibility
 ($\chi^{(2)}$) without electric-field poling.
\end{tocentry}

\begin{abstract}

Materials with large second-order nonlinearities are crucial for next-generation integrated photonics. Spontaneously oriented organic thin films prepared by physical vapor deposition offer a promising poling-free and scalable approach. This study investigates molecular engineering strategies to enhance the second-order nonlinear response of derivatives based on the donor-acceptor molecule 2-(4'-diphenylaminobiphenyl-4-yl)quinoxaline-6,7-dicarbonitrile (TPA-QCN). Four derivatives incorporating modifications designed to increase molecular hyperpolarizability ($\beta$) or promote favorable orientation were synthesized and characterized. The most successful modification, intramolecular bridge-locking, simultaneously increases hyperpolarizability and enhances spontaneous orientation by reducing detrimental electrostatic interactions during deposition. It leads to a significant enhancement of the second-order nonlinear response, achieving off-resonance $\chi^{(2)}_{31} \approx 16$ pm V$^{-1}$ and $\chi^{(2)}_{33} \approx 18$~pm V$^{-1}$ at 1550~nm, a twofold improvement over the parent TPA-QCN. Analysis combining nonlinear optical measurements, surface potential measurement, optical anisotropy, and density functional theory  calculations indicates that improved molecular orientation, rather than increased $\beta$ alone, is the primary driver for the enhanced performance in the leading derivatives. These findings demonstrate the effectiveness of targeting molecular orientation via structural design and position spontaneously oriented organic films as compelling  poling-free candidates for integrated nonlinear photonic applications where the increased  electrode-induced optical losses, fabrication complexity and footprint are a critical limitation.

\end{abstract}


\section{Introduction}

The development of advanced photonic technologies, with the potential to transform areas like high-speed communication\cite{Dalton_Leuthold_Robinson_Haffner_Elder_Johnson_Hammond_Heni_Hosessbacher_Baeuerle_etal_2023}, quantum computing\cite{zhong2020quantum,o2007optical, madsen2022quantum}, and advanced sensing\cite{Lawrie_Lett_Marino_Pooser_2019, Polino_Valeri_Spagnolo_Sciarrino_2020, Peters_Rodriguez_2022}, is intrinsically linked to the ability to engineer strong optical nonlinearities\cite{hendrickson2014integrated}. Second-order nonlinearities, which typically surpass third-order nonlinearities in strength, enable key nonlinear processes like electro-optic modulation, optical parametric oscillation and sum-frequency generation \cite{boyd2020nonlinear}. However, realizing these benefits in practical devices requires materials that combine large nonlinearities and  compatibility with established nanofabrication techniques used in large-scale fabrication, a challenging combination. While lithium niobate \cite{Boes_Chang_Langrock_Yu_Zhang_Lin_Lončar_Fejer_Bowers_Mitchell_2023} and poled electro-optic organic molecules \cite{Dalton_Leuthold_Robinson_Haffner_Elder_Johnson_Hammond_Heni_Hosessbacher_Baeuerle_etal_2023} have promising nonlinear properties, they face significant limitations. Lithium niobate has a large second-order optical nonlinearity but is difficult to integrate with CMOS silicon photonics due to lithium contamination and specialized etching needs \cite{tan2024micro}. Conversely, organic molecules can possess exceptionally large microscopic nonlinearities (hyperpolarizabilities, $\beta$)\cite{Nalwa_Miyata_1997,Biaggio_2022}, but typically assemble centrosymmetrically in films when processed from solution \cite{Dalton_Sullivan_Bale_2010}or the vapor phase, thus resulting in films lacking bulk $\chi^{(2)}$ activity. 
The required non-centrosymmetry is typically induced via electric-field poling.\cite{Wu_Li_Luo_Jen_2020} This technique is highly effective and has led to electro-optic materials with exceptionally large macroscopic nonlinearities and electro-optic response, far exceeding those of inorganic crystals like lithium niobate. However, poling introduces significant practical challenges for device integration. It requires electrodes, which not only add process complexity for back-end-of-the-line (BEOL) fabrication \cite{Taghavi_Moridsadat_Tofini_Raza_Jaeger_Chrostowski_Shastri_Shekhar_2022} but also introduce optical absorption losses. These losses are particularly detrimental for applications like quantum light generation or low-threshold optical parametric oscillation where minimizing optical loss is a primary concern\cite{Dutt_Mohanty_Gaeta_Lipson_2024} and where electrodes are not otherwise required. While the losses can be mitigated by increasing the electrode-waveguide separation, this comes at the direct cost of a larger device footprint. A monolithically integrated, poling-free material with a significant $\chi^{(2)}$ would therefore be highly advantageous for such loss-sensitive applications, as it also reduces the fabrication complexity and the device footprint.

\begin{figure}[H]
    \centering
    \includegraphics[width=0.8\linewidth]{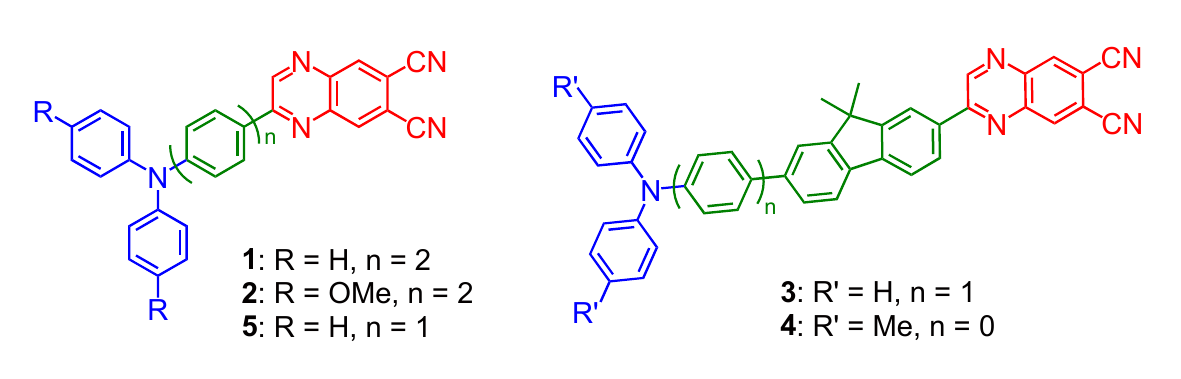}
    \caption{Investigated donor(D)-acceptor(A) molecules in the study. D group is in blue, A group is in red, $\pi$-linker is in green.}
    \label{Fig0:Compounds}
\end{figure}

Physical vapor deposition (PVD) of specifically designed organic molecules offers such a pathway through \textit{spontaneous orientation}\cite{Theriault_et_al}. During vacuum thermal evaporation, a complex interplay between intermolecular forces (\textit{e.g.}, dipole-dipole interactions), molecular packing, and surface energy can lead to a  non-centrosymmetric out-of-plane orientation distribution of molecules \cite{Hofmann_Schmid_Brütting_2021,Pakhomenko_He_Holmes_2023,Tanaka_Auffray_Nakanotani_Adachi_2022}. This spontaneously broken inversion symmetry enables bulk $\chi^{(2)}$ effects without external poling. Furthermore, PVD is compatible with large-scale CMOS fabrication, yielding highly uniform thin films\cite{leuthold2009silicon, Biaggio_2022}. We recently demonstrated this potential using the donor-acceptor (D-A) molecule TPA-QCN (compound \textbf{1}, Figure~\ref{Fig0:Compounds}), achieving promising second-order susceptibilities ($\chi_{31}^{(2)}=15\pm1$~pm V$^{-1}$ and $\chi_{33}^{(2)}=14\pm4$~pm V$^{-1}$ at 1266~nm) in neat evaporated films\cite{Theriault_et_al}.

To enhance these nonlinearities, further molecular engineering is required. The macroscopic $\chi^{(2)}$ susceptibility depends on both the molecular hyperpolarizability ($\beta$) and the degree of collective non-centrosymmetric orientation\cite{Singer_Kuzyk_Sohn_1987}. Strategies must therefore target one or both factors. While designing molecules for high $\beta$ is relatively well-understood\cite{Dalton_Sullivan_Bale_2010,Marder_Kippelen_Jen_Peyghambarian_1997, Wu_Li_Luo_Jen_2020,Barlow_Marder_2006}, predicting and controlling spontaneous orientation during PVD remains challenging, relying largely on empirical observations related to molecular shape and deposition conditions. Moreover, modifications aimed at increasing $\beta$ can inadvertently disrupt favorable orientation, potentially reducing the overall $\chi^{(2)}$. Addressing this challenge, we synthesized and investigated four TPA-QCN derivatives (Figure~\ref{Fig0:Compounds}) to explore the following structural effects:
\begin{itemize}
    \item Modifying the donor with strong methoxy groups to increase hyperpolarizability\cite{Dalton_Sullivan_Bale_2010,wu2014towards,suresh2005synthesis,muhammad2016impact,ward2019impact,wazzan2019effect}. (compound \textbf{2})
    \item Lengthening the $\pi$-bridge to enhance hyperpolarizability\cite{blanchard1995large, Barlow_Marder_2006} (compound \textbf{3})
    
    \item Increasing molecular planarity through bridge locking to increase hyperpolarizability\cite{Li_Xiao_Shen_Deng_LongGu_2022} (compound \textbf{4})
    
    \item Removing the $\pi$-bridge between the donor and acceptor to promote spontaneous orientation\cite{Yokoyama_2011} and better conjugation (compound \textbf{5}).
\end{itemize}

This work details the design rationale, synthesis, computational validation, and comprehensive optical characterization of these derivatives, providing insights into the structure-property relationships governing second-order nonlinearity in spontaneously oriented films.

\section{Design Rationale and Computational Validation}
\begin{figure}[H]
    \centering
    \includegraphics[width=0.8\linewidth]{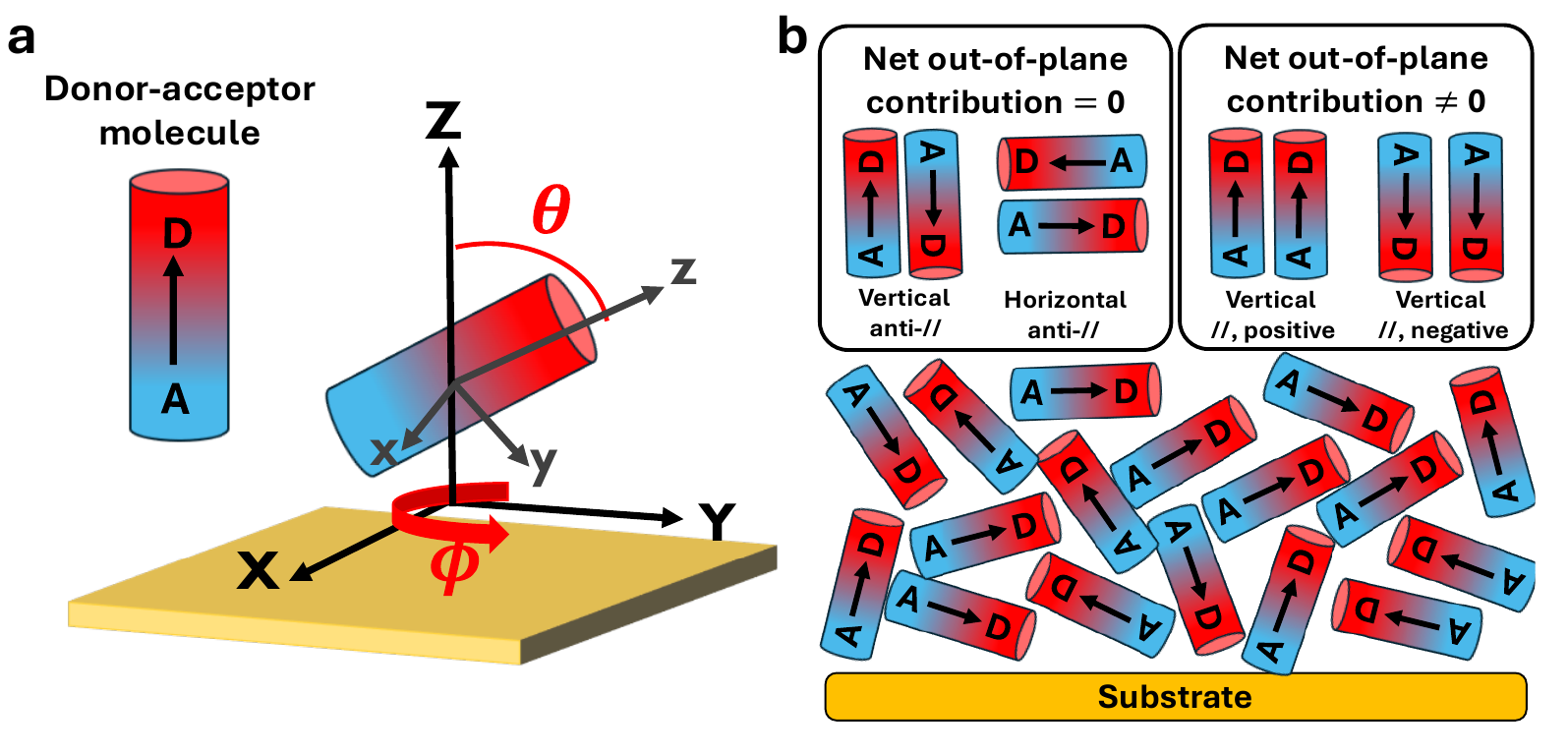} 
    \caption{(a) Schematic of a D-A molecule on a substrate defining the tilt angle ($\theta$) and azimuthal angle ($\phi$). The dependence on the azimuthal angle vanishes due to isotropy of the in-plane direction \cite{Yokoyama_2011}. The black arrow represents the direction of the permanent dipole moment, from a negatively-charged region to a positively-charged region. (b) Pictorial representation
of a thin film with D-A molecules and how molecular packing configurations can lead to the
presence or absence of a net out-of-plane contribution.  In the configuration with zero net
contribution, the molecules are arranged in a balanced anti-parallel fashion.  A purely horizontal parallel configuration is not observed macroscopically. Due to the in-plane rotational symmetry of the deposition (an amorphous substrate, a homogeneous in-plane environment, and substrate rotation), there is no preferred in-plane direction. Therefore, the in-plane components must average to zero over a macroscopic area, resulting in a film with overall $C_{\infty v}$ symmetry.\cite{Yokoyama_2011}. In the non-zero contribution scenario, parallel alignment leads
to a net out-of-plane contribution(required for $\chi^{(2)}$).The net out-of-plane contribution in the schematic of the film (bottom) is exaggerated to clearly illustrate a non-centrosymmetric arrangement.}
    \label{Fig1}
\end{figure}

Spontaneously oriented polar films grown by PVD  exhibit $C_{\infty,v}$ symmetry, possessing two independent non-zero $\chi^{(2)}$ tensor elements: $\chi^{(2)}_{33}$  and $\chi^{(2)}_{31}$ \cite{Popov_Svirko_Želudev_2017}. For rod-like D-A molecules like \textbf{1} (see  Supporting Information  S1 for discussion about its structure), where the primary component of the hyperpolarizability tensor ($\beta_{zzz}$) lies along the long molecular axis, these macroscopic susceptibilities are related to $\beta_{zzz}$ and ensemble averages of the molecular orientation angle $\theta$ (relative to the surface normal, see Figure~\ref{Fig1}a) by:\cite{Singer_Kuzyk_Sohn_1987}
\begin{align}
    \chi^{(2)}_{33} &\propto N \beta_{zzz} \langle \cos^3\theta \rangle \\
    \chi^{(2)}_{31} &\propto N \beta_{zzz} \frac{1}{2}(\langle \cos\theta\rangle - \langle\cos^3\theta \rangle)
\end{align}

where $N$ is the molecular number density. Achieving large $\chi^{(2)}$ values thus requires maximizing both  the product of $N$ and $\beta_{zzz}$ and the relevant orientational average terms, which reflect the degree of out-of-plane non-centrosymmetric orientation. For many rod-like molecules, the orientation distribution in spontaneously formed PVD films is predominantly horizontal (i.e., with $\theta$ approaching 90°)\cite{Yokoyama_2011}. The net macroscopic nonlinearity arises from a small but non-negligible polar asymmetry within this predominantly horizontal arrangement. Figure~\ref{Fig1}b schematically illustrates this concept: while configurations like balanced anti-parallel vertical pairs (top left panel) lead to cancellation of molecular dipoles and zero net $\chi^{(2)}$, parallel vertical configurations result in a  non-centrosymmetric out-of-plane distribution (\textit{e.g.}, more molecules pointing 'up' than 'down', bottom right panel) that yields the required macroscopic nonlinearity.

The precise mechanisms driving spontaneous orientation are complex, involving competing factors like dipole-dipole interactions (favoring anti-parallel pairing, reducing net asymmetry), surface energy minimization (favoring parallel alignment), and molecular shape anisotropy (with linear molecules tending towards horizontal packing, and compact molecules potentially favoring more vertical orientations)\cite{Tanaka_Auffray_Nakanotani_Adachi_2022, Pakhomenko_He_Holmes_2023,Hofmann_Schmid_Brütting_2021,Yokoyama_2011}. The lack of clear predictive rules necessitates exploring different molecular architectures to identify designs that effectively promote spontaneous orientation. Alternative approaches like supramolecular assembly have also been explored to induce non-centrosymmetry without poling\cite{Facchetti_Annoni_Beverina_Morone_Zhu_Marks_Pagani_2004, Zhu_Kang_Facchetti_Evmenenko_Dutta_Marks_2003, Muller_Cai_Kündig_Tao_Bösch_Jäger_Bosshard_Günter_1999}. However, these methods rely on highly specific intermolecular interactions (e.g., hydrogen bonding), which can impose significant constraints on molecular design  and thermal stability of the chromophores compared to the broader applicability of intrinsic spontaneous orientation in evaporated films. Our reference molecule, \textbf{1}, possesses a near-linear D-$\pi$-A structure and was previously predicted\cite{Li_Xiao_Shen_Deng_LongGu_2022}  to have a large static hyperpolarizability  in a computational study using an optimally tuned range-separated $\omega$B97XD functional.

To validate the impact of molecular modifications on the properties of the proposed compounds, we first calculated their molecular hyperpolarizabilities and ground state dipole moment using density functional theory (DFT, Table~\ref{tbl:properties}) \cite{Lescos_Sitkiewicz_Beaujean_Blanchard-Desce_Champagne_Matito_Castet_2020}. All molecules adopt a similar rod-like shape with the length varying from 1.98~nm (\textbf{5}) to 2.80~nm (\textbf{3}). Notably, the longest molecule~\textbf{3} shows the lowest ground state dipole moment indicating the reduced polarization across the three benzene-rings long bridge. On the other hand, the electron-donating substituents on the D termini increase the dipole moment (\textit{cf.} \textbf{2} vs \textbf{1}).For the analysis, the molecular coordinate system was defined with the z-axis along the ground-state dipole moment. In this frame, the hyperpolarizability tensor of all compounds (Supporting Information  S2, Table S3) is strongly dominated by the $\beta_{zzz}$ component. Increasing the donor ability of the TPA moiety with methoxy substituents significantly increases the hyperpolarizability in chromophore~\textbf{2}. Increasing the length of the $\pi$-bridge in NLO chromophores significantly enhances their $\beta$ values as observed before\cite{moore2001synthesis,marder1994large}. In our series \textbf{1}--\textbf{5} we observe that the $\beta$ is maximized for $\pi$-bridges consisting of two benzene rings and further elongation does not lead to higher hyperpolarizability (\textit{cf.}~\textbf{3} vs~\textbf{4}), similar what was observed in other D/A substituted oligophenylenes\cite{cheng1991experimental}. This limited effective conjugation is due to high aromaticity as well as the inter-ring twist of the oligophenylene bridge. Indeed, the calculated $\beta_{\text{zzz},0}$ of \textbf{1} is $145 \times 10^{-30}$~esu in its lowest energy conformation (Ph-Ph torsion angle 32\textdegree). This value increases to $211 \times 10^{-30}$~esu for the co-planar (0\textdegree) conformation and drops to as low as $42 \times 10^{-30}$~esu in the orthogonal (90\textdegree) conformation (Figure~S6). Introduction of the fluorene linker locks the Ph-Ph interring twist resulting in a 34\% hyperpolarizability increase in~\textbf{4} compared to its biphenylene homologue~\textbf{1}. Finally, it is also crucial to consider the molecular number density ($N$), as the intrinsic molecular contribution to the film's nonlinearity is best represented by the $N|\beta_{zzz,0}|$ figure of merit (Table \ref{tbl:properties}). This metric reveals a different perspective from $\beta_{zzz,0}$ alone. The case of compound \textbf{5} is particularly instructive: its relatively low hyperpolarizability is partially compensated by its high number density, which results from its compact structure. This highlights a key design principle: optimizing $\beta_{zzz,0}$ is insufficient, as structural modifications aimed at increasing hyperpolarizability must not come at the expense of a significant decrease in number density.

\begin{table}[H]
  \caption{Static hyperpolarizabilities\cite{Lu_Chen_2012} and dipole moment calculated in gas phase for the compounds \textbf{1}--\textbf{5} in their lowest energy conformations with M062x functional and 6-31G(d)  basis set, number density (see Supporting Information S3 for details) and the $N|\beta_{zzz,0}|$ figure of merit.  }
  \label{tbl:properties}
  \footnotesize
  \begin{tabular}{cccccc}
    \hline
Compound  & $\beta_{zzz,0}$ & Dipole moment&Number density $N$ & $N|\beta_{zzz,0}|$\\
            &[$10^{-30}$ esu] &    [D] &[$10^{27}\text{m}^{-3}$]&[ esu $\text{m}^{-3}$ ]\\
    \hline
   \textbf{1}    &-145&10.6&1.54&0.22\\
    \textbf{2} &-192  &12.3 &1.31&0.25\\   
    \textbf{3} &-159  &10.5&1.40&0.22\\
    \textbf{4} &-195 &11.2 &1.42&0.27 \\
    \textbf{5}&-129 &11.2 &1.89&0.24\\
    \hline
  \end{tabular}
\end{table}

\section{Results}

The purified materials were thermally evaporated onto fused silica and thermally oxidized silicon substrates in a vacuum chamber(see Methods for details), yielding films with thicknesses ranging from 49 to 108 nm, depending on the material availability. The morphology of the evaporated films was first characterized using X-ray diffraction (XRD). The diffractograms for all compounds show a single broad featureless peak originating from the substrate, confirming that the films are amorphous (see Supporting Information S4, Figure S3). As a second characterization step, the linear optical properties were characterized by measuring absorbance spectra and performing variable-angle spectroscopic ellipsometry (see Methods and Supporting Information S5 for details).

\begin{figure}[H]
    \centering
    \includegraphics[width=0.45\linewidth]{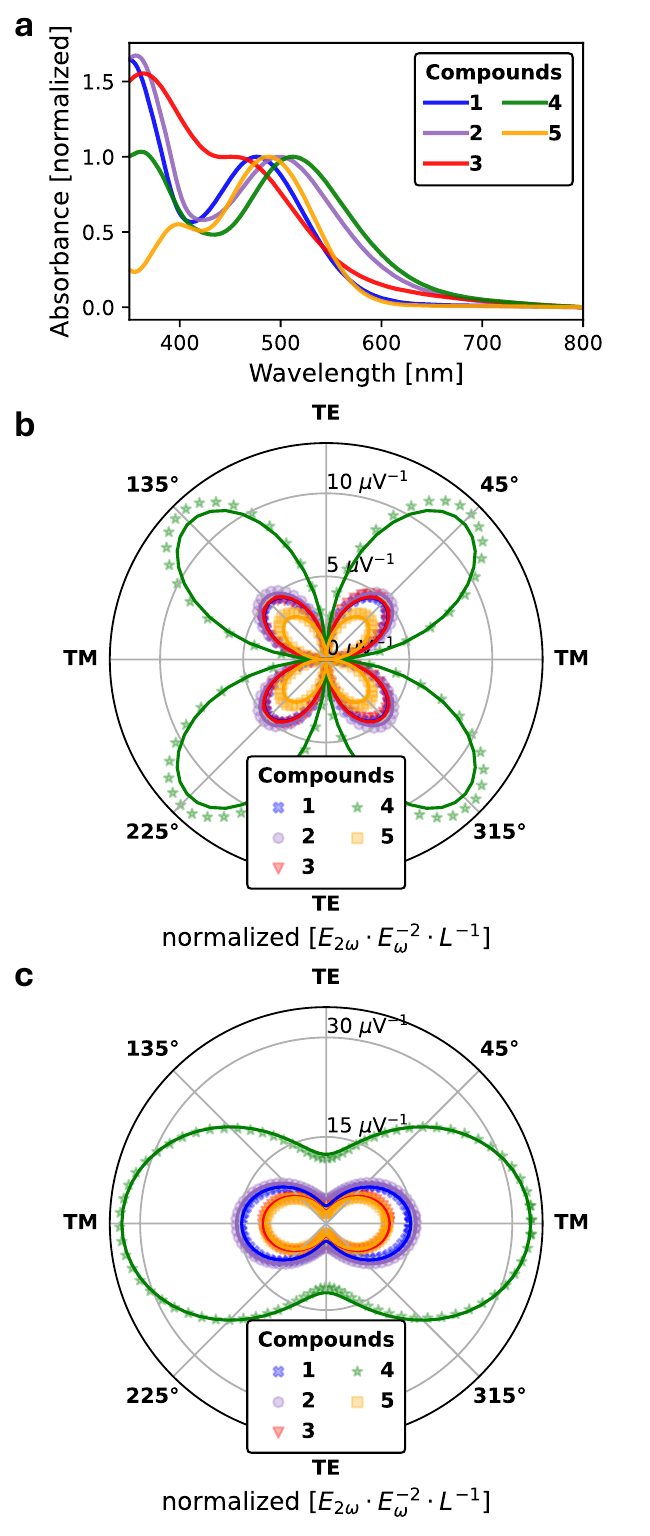} 
    \caption{Linear and nonlinear optical properties. (a) Normalized absorption spectra (b, c) SHG polarimetry data (TE and TM pump) at $\lambda_{pump}=1550$~nm, incidence angle 45$^\circ$. Field strengths normalized by $E_{inc}^2 \times L$. Peak intensities: 30-100 GW/cm$^2$. Points: data; Lines: fits from nonlinear transfer matrix. On panel (b), data and fits from compounds \textbf{1}, \textbf{2} and \textbf{3} are overlapping. On panel (c), data and fits from compounds \textbf{3} and \textbf{5} are overlapping. }
    \label{Fig2}
\end{figure}

Figure \ref{Fig2}a shows the normalized linear absorbance spectra of the different films. All films exhibit a charge-transfer absorption peak centered between 450 and 515 nm. With the exception of \textbf{3}, which has a blue-shifted peak (451 nm) relative to \textbf{1} (476 nm), all other compounds are red-shifted. \textbf{4} exhibits the largest red-shift, with absorption centered at 512 nm. \textbf{2} and \textbf{4} have absorption tails extending significantly further into the visible (up to ~720 nm), while \textbf{3} shows tails up to ~700 nm, and \textbf{5} and \textbf{1} show minimal absorption above 620 nm. Time-dependent DFT (TDDFT, Supporting Information S2) calculations attribute these absorption bands to the allowed intramolecular charge transfer transitions (S0-S1) and reproduce the observed trend, although the calculated absorptions were blue-shifted compared to the experimentally measured spectra by 0.38-0.42 eV. Furthermore, transmittance measurements confirmed that all films are  transparent in the near-infrared, with no significant absorption features observed out to 2500 nm (see Supporting Information S6, Figure S10-S14). Spectroscopic ellipsometry measurements were used to extract the refractive indices of the different films (Supporting Information S5).

Next, second-order nonlinear properties were characterized using second-harmonic polarimetry (see Methods for details) with two different pump wavelengths: $\lambda$ = 1266 nm and $\lambda$ = 1550 nm. At 1266 nm, the second-harmonic wavelength (633 nm) is near resonance with the absorption tails of some materials (\textbf{2}, \textbf{3}, \textbf{4}),  which is expected to lead to significant  resonant enhancement of the $\chi^{(2)}$ response. At 1550 nm (second-harmonic at 775 nm), resonant effects are expected to play a lesser role.

Figures \ref{Fig2}b and \ref{Fig2}c show the transverse electric (TE) and transverse magnetic (TM) second-harmonic polarimetry data for the different films at 1550 nm pump wavelength. Data are presented as second-harmonic field strength normalized by incident field squared and thickness for easier comparison. \textbf{4} stands out with a field strength approximately twice as strong as the next highest signal (from \textbf{1} and \textbf{2}).

\begin{figure}[H]
    \centering
    \includegraphics[width=0.9\linewidth]{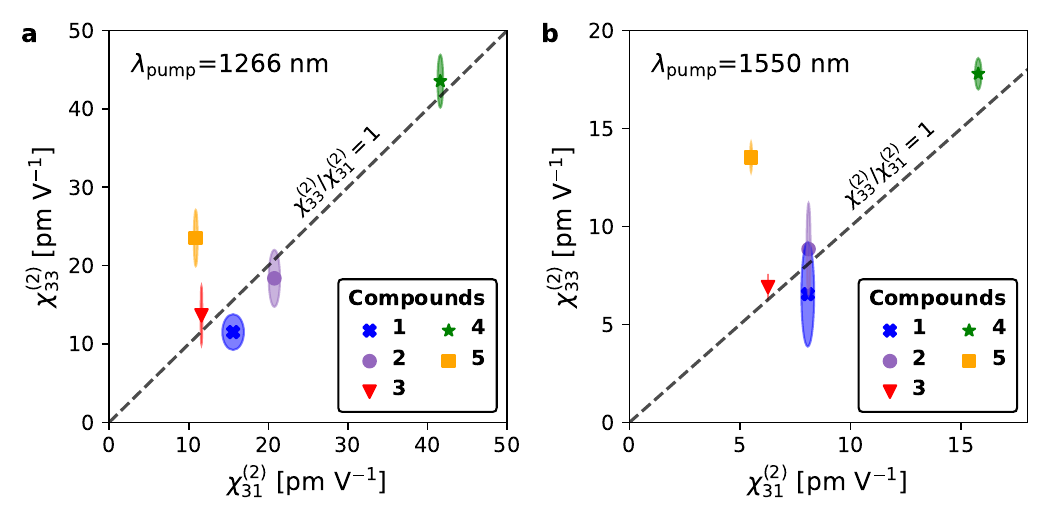} 
    \caption{Measured second-order susceptibilities at (a) $\lambda_\text{pump}=1266$~nm and (b) $\lambda_\text{pump}=1550$~nm. Plots show $\chi^{(2)}_{33}$ vs. $\chi^{(2)}_{31}$. Points and shaded regions represent weighted means and uncertainties. Dashed line: $\chi^{(2)}_{33}/\chi^{(2)}_{31} = 1$. Note the significant enhancement for \textbf{4}.}
    \label{Fig3}
\end{figure}

Nonlinear coefficients were extracted using a nonlinear transfer matrix algorithm\cite{Bethune_1991} (see Methods). Simulated patterns are shown as solid lines in Figure \ref{Fig2}b and \ref{Fig2}c and are in excellent agreement with the data. The simulation accounts for material dispersion and anisotropy (obtained from ellipsometry), interference effects, and absorption losses. Measurements were performed at a minimum of two incidence angles per film. Reported coefficients are uncertainty-weighted means of extracted coefficients at different angles. Results are shown in Figures \ref{Fig3}a ($\lambda_{pump}$ = 1266 nm) and \ref{Fig3}b ($\lambda_{pump}$ = 1550 nm).

As anticipated from the polarimetry data, \textbf{4} exhibits remarkably high second-order nonlinear coefficients for both $\chi^{(2)}_{31}$ and $\chi^{(2)}_{33}$, exceeding 40 pm V$^{-1}$ at 1266 nm—a threefold improvement over \textbf{1}. At 1550 nm, the enhancement for \textbf{4} over \textbf{1} is approximately twofold. The larger enhancement at 1266 nm suggests a contribution from resonant effects, which are expected to be stronger for the more red-shifted compound \textbf{4}. Nevertheless, the off-resonance values at 1550 nm for \textbf{4} ($\chi^{(2)}_{31} \approx 16$ pm V$^{-1}$ and $\chi^{(2)}_{33} \approx 18$~pm V$^{-1}$) are notable, with $\chi^{(2)}_{31}$ significantly exceeding values reported for common nonlinear integrated photonics materials like aluminum nitride ($\chi^{(2)}_{31} \sim 0.2$ pm V$^{-1}$, $\chi^{(2)}_{33} \sim 9$~pm V$^{-1}$ at 1030 nm)\cite{Majkić_Franke_Kirste_Schlesser_Collazo_Sitar_Zgonik_2017} and lithium niobate ($\chi^{(2)}_{31} \sim 6$~pm V$^{-1}$, $\chi^{(2)}_{33} \sim 40$~pm V$^{-1}$ at 1310 nm)\cite{Shoji_Kondo_Kitamoto_Shirane_Ito_1997}. The $\chi^{(2)}_{31}$ element is particularly relevant for type-I or type-II interactions, such as polarization-entangled photon generation\cite{kang2016monolithic,zhang2024polarization}.

The other derivatives showed smaller changes compared to TPA-QCN (\textbf{1}). \textbf{2} shows slight enhancement at 1266 nm but nearly identical nonlinear properties at 1550 nm, consistent with resonance enhancement playing a role at the shorter wavelength due to its red-shifted absorption. \textbf{3} has similar $\chi^{(2)}_{33}$ but $\sim$25\% lower $\chi^{(2)}_{31}$ than \textbf{1}. \textbf{5}, despite having a significantly lower computed $\beta_{zzz,0}$ and exhibiting 25-30\% lower $\chi^{(2)}_{31}$ values, shows a nearly twofold enhancement in $\chi^{(2)}_{33}$ compared to TPA-QCN. Notably, \textbf{5} is the only compound where $\chi^{(2)}_{33}$ is significantly larger than $\chi^{(2)}_{31}$, suggesting a distinct average molecular orientation distribution compared to the other derivatives.

\section{Discussion}

While compound \textbf{4} clearly exhibits the highest overall second-order nonlinearity among the studied derivatives, understanding the origin of this enhancement—whether driven by molecular hyperpolarizability ($\beta$) or superior spontaneous orientation is crucial for guiding future molecular engineering effort. To disentangle these contributions, we analyzed the nonlinear coefficients normalized by the computed static hyperpolarizability component along the molecular axis, the number density and the resonant enhancement factor($\chi^{(2)}_{ij}/N F\beta_{zzz,0}$).The calculated resonant enhancement factors ($F$) \cite{Singer_Kuzyk_Sohn_1987} using a two-level model are shown in  Table S7 (Supporting Information S7). The values confirm that, relative to the reference compound \textbf{1}, while some enhancement is still present at 1550 nm, it is considerably lower than at 1266 nm for all compounds. This normalization factor isolates the impact of molecular orientation; if performance were solely dictated by the product of $\beta$, $N$ and $F$ , this metric would be constant across all compounds.

It is important to note that the $\beta_{zzz,0}$ values used for this normalization are calculated for the gas phase with the M062x functional. This is an approximation, as the hyperpolarizability is sensitive to both the local dielectric environment, the molecular conformation (which may be different from the gas phase predictions) and the choice of DFT functional. However, as our comparative study in the Supporting Information (Table S1) demonstrates, while these factors can significantly alter the absolute hyperpolarizability values, the key relative trends between the derivatives remain consistent, and the central conclusions drawn below are not affected.

\begin{figure}[H]
    \centering
    \includegraphics[width=0.9\linewidth]{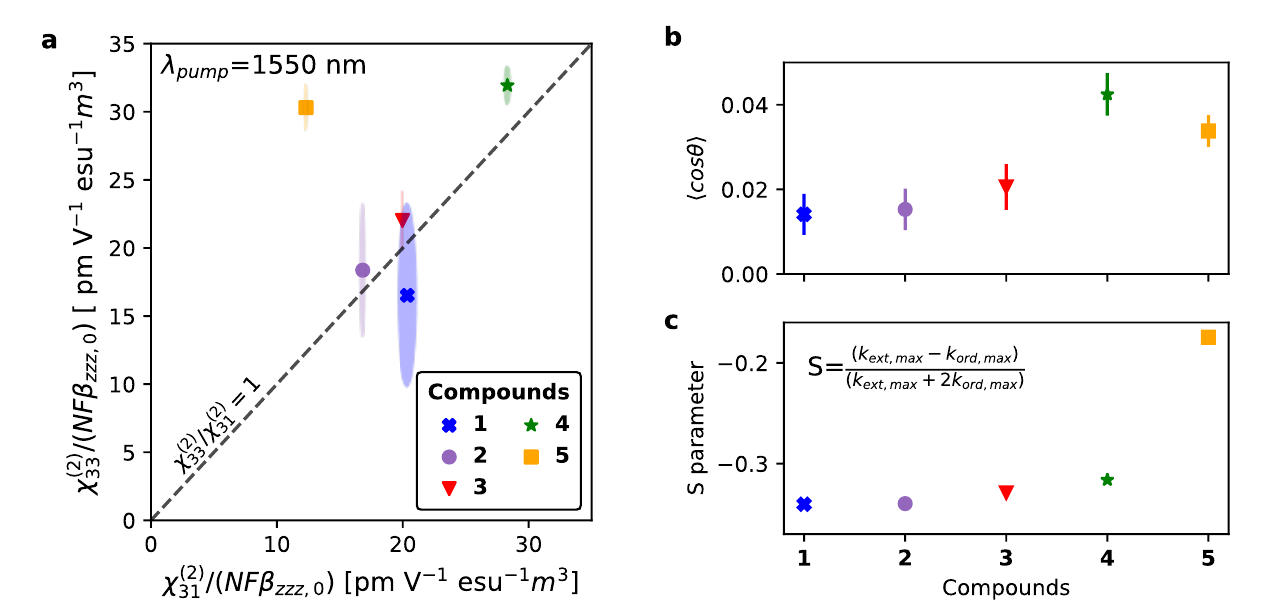}
\caption{(a)  Second-order nonlinear susceptibility normalized by both the molecular hyperpolarizability, number density and field enhancement factor ($\chi^{(2)}_{ij}/(NF\beta_{zzz,0})$). The dashed line indicates $\chi^{(2)}_{33}/\chi^{(2)}_{31} = 1$. (b)  First-order orientational moment, $\langle cos \theta \rangle$, calculated from the measured surface potential and number density (see Supporting Information S8).(c) Optical anisotropy parameter S, derived from the peak values of the unaxial imaginary refractive index components ($k_{\text{ord,max}}, k_{\text{ext,max}}$ as defined in the panel equation) and reflecting the average TDM orientation (S=0: isotropic, S=-0.5: horizontal, S=1: vertical).}
    \label{Fig4}
\end{figure}

The normalized coefficients, presented in Figure~\ref{Fig4}a, reveal that orientation effects play a dominant role. \textbf{4}'s superior performance persists even after accounting for its 22\% larger $N\beta_{zzz,0}$ and 14\% larger  resonant enhancement, indicating that its enhanced $\chi^{(2)}$ is driven predominantly by a more favorable molecular alignment during film growth. Similarly,  \textbf{5}, despite possessing the lowest $\chi^{(2)}_{31}$ and a $\sim$50\% lower computed $\beta_{zzz,0}$, emerges as a top performer in this normalized comparison for the $\chi^{(2)}_{33}$ component. This strongly suggests a highly effective orientational distribution for maximizing $\chi^{(2)}_{33}$ in this molecule. Conversely, \textbf{2}, which boasts the second highest computed molecular hyperpolarizability, shows relatively poor performance in the normalized metric, pointing towards a less advantageous spontaneous orientation distribution that counteracts its high $\beta$.

Further evidence supporting the importance of orientation comes from surface potential measurements, which probe the first order orientational moment $\langle cos \theta \rangle$. The results, presented on  Figure~\ref{Fig4}b as the first-order orientational moment are derived from the Giant Surface Potential (GSP) slope (surface potential normalized by film thickness). Additional details about this measurement are available in the Methods section and Supplementary Information S8. Consistent with the normalized $\chi^{(2)}$ analysis, compounds \textbf{4} and \textbf{5} show significantly larger $\langle cos \theta \rangle$ (around 0.035-0.040) than the other derivatives (\textbf{1}, \textbf{2}, \textbf{3},  with values around 0.02), confirming a stronger net spontaneous polarization in \textbf{4} and \textbf{5}. The positive sign of the $\langle cos \theta \rangle$ for all compounds indicates a preferential orientation with the negatively-polarized QCN group pointing towards the substrate, on average. However, the small magnitude of these values indicates that the overall net polar order is weak.

While this analysis highlights the importance of orientation for compounds \textbf{4} and \textbf{5}, it also reveals differences in their alignment characteristics. Compound \textbf{5} uniquely displays a $\chi^{(2)}_{33}/\chi^{(2)}_{31}$ ratio significantly greater than unity ($\approx 2$), unlike the other derivatives where this ratio is closer to 1. This suggests a different out-of-plane angle ($\theta$) distribution for \textbf{5}. Insights into this distribution can be gained from the optical anisotropy parameter $S$, which probes the orientation of the transition dipole moment (TDM)\cite{Hofmann_Schmid_Brütting_2021}.Crucially, for quasi-linear D-$\pi$-A molecules like those studied here, the TDM, the permanent dipole moment, and the molecular long axis (relevant for $\beta_{zzz}$) are approximately collinear (Supporting Information S2). The S parameter, shown in Figure~\ref{Fig4}c, is derived from the anisotropy in the  imaginary part of the refractive index (Supporting Information S5) and is related to an even-order moment of the orientation distribution ($\langle \cos^2\theta \rangle$). It thus describes the degree of molecular alignment irrespective of the polarity (proportional to odd moments)  relevant to $\chi^{(2)}$ measurements.  While all films show $S < 0$, indicating a preference for horizontal alignment (molecular axis parallel to the substrate), compound \textbf{5} exhibits a value ($S \approx -0.18$, corresponding to $\langle cos^2\theta\rangle\approx0.22$) significantly closer to isotropy ($S=0$, $\langle cos^2\theta\rangle=0.33$) compared to the others ($S \approx -0.30$ to $-0.35$, $\langle cos^2\theta\rangle\approx0.10-0.12$). This indicates its orientation distribution is  less biased towards horizontal alignment compared to the other derivatives, likely influenced by its compact structure\cite{Yokoyama_2011}. 

The ratio $\chi_{33}^{(2)}/\chi_{31}^{(2)}$ is highly sensitive to the specific form of the molecular orientation distribution, as $\chi_{33}^{(2)}$ depends directly on the third-order moment $\langle \cos^3\theta \rangle$, while $\chi_{31}^{(2)}$ involves $\frac{1}{2}(\langle \cos\theta \rangle - \langle \cos^3\theta \rangle)$. For compound \textbf{5}, GSP and $\chi^{(2)}$ measurements confirm a net polar order (non-zero $\langle \cos\theta \rangle$ and $\langle \cos^3\theta \rangle$). The optical anisotropy parameter S, which reflects $\langle \cos^2\theta \rangle$, further reveals that compound \textbf{5}'s alignment is notably less confined to the horizontal plane than other derivatives. This suggests that its orientation distribution is broader, with a greater population in what can be considered the 'tails' of the distribution (the more upright orientations where  $\theta$ approaches 0 or $\pi$).  Indeed, such vertical configurations correspond to high magnitudes of $\cos\theta$. Higher-order moments like $\langle \cos^3\theta \rangle$ are inherently more sensitive to these tails of the distribution, where the $\cos^3\theta$ term itself is maximized. Consequently, an increased asymmetry in the tails can disproportionately boost $\langle \cos^3\theta \rangle$ relative to changes in $\langle \cos\theta \rangle$, providing a plausible explanation for the larger $\chi_{33}^{(2)}/\chi_{31}^{(2)}$ ratio observed for compound \textbf{5}.

Explaining the superior orientation achieved by compound \textbf{4}, despite its increased planarity typically favoring horizontal alignment\cite{Yokoyama_2011}, requires further consideration. We hypothesize that the rigid, locked fluorene-containing bridge structure plays a crucial role. First, the locked bridge enhances the molecular hyperpolarizability  (22\% increase in $N \beta_{zzz,0}$), and second, its more red-shifted absorption leads to a slightly larger resonant enhancement (14\% increase for $\lambda_p=1550$ nm) when compared to \textbf{1}. However, these two factors combined are not sufficient to explain the observed twofold increase in $\chi^{(2)}$. Therefore, we propose that the steric hindrance provided by the C(CH$_3$)$_2$ groups on the fluorene linker plays an important role by frustrating   the close anti-parallel packing driven by dipole-dipole interactions. This hypothesis is directly analogous to a well-established strategy in electrically poled systems, where such steric effects are known to reduce detrimental dipole pairing and improve a efficiency\cite{shi2000low,Dalton_Sullivan_Bale_2010,liu2019synthesis,dalton1997role}. In the context of spontaneous orientation, reducing the formation of centrosymmetric anti-parallel pairs is key. Such pairs lead to cancellation in the orientation averages $\langle \cos\theta \rangle$ and $\langle \cos^3\theta \rangle$ that determine $\chi^{(2)}$. Therefore, minimizing these pairs results in a larger net polar alignment and, consequently, a higher macroscopic $\chi^{(2)}$, consistent with our observations for \textbf{4} (large $\chi^{(2)}$, high $\langle cos \theta\rangle $ and low S values). This contrasts with compound \textbf{3}, which also incorporates a fluorene unit but via a longer, unlocked bridge. We propose that for \textbf{3}, any potential benefit from reduced dipole interactions is likely offset by the increased tendency of longer, linear molecules to align horizontally\cite{Yokoyama_2011}, thus diminishing the overall $\chi^{(2)}$ response, a fact consistent with the low S value and GSP slope.

These findings highlight two distinct and effective molecular engineering strategies for enhancing $\chi^{(2)}$ in spontaneously oriented films: first, achieving a molecular orientation distribution less dominated by purely horizontal alignment, and second, directly reducing detrimental anti-parallel dipole pairing. Compound \textbf{5} exemplifies the first strategy; its more compact structure, arising from the removal of a $\pi$-bridge, leads to an alignment that is less restricted to the horizontal plane\cite{Yokoyama_2011} (as indicated by its comparatively larger S parameter, signifying a weaker preference for horizontal orientation than other derivatives). This altered distribution appears to favorably impact the critical orientation averages governing $\chi^{(2)}$. For compound \textbf{4}, the second strategy is prominent: the rigid, locked fluorene bridge appears to effectively frustrate anti-parallel packing. This increases the net polar order and boosts the macroscopic $\chi^{(2)}$, even though compound \textbf{4} maintains a significant  preference for horizontal molecular alignment. Ultimately, these results underscore that successful molecular engineering requires a delicate balance between intrinsically coupled design strategies. Modifications aimed at promoting net polar order, either by reducing anti-parallel pairing or by encouraging a more vertical orientation, must be implemented without sacrificing the intrinsic molecular nonlinearity, and more specifically the $N\beta_{zzz,0}$
figure of merit.

Finally, translating these promising material properties into practical device applications necessitates addressing factors like thermal stability. The glass transition temperature ($T_g$)  typically dictates the operational range and compatibility with fabrication processes, as spontaneous orientation is \st{typically} anticipated to be lost above $T_g$ \cite{Pakhomenko_He_Holmes_2023}. The $T_g$ of the different derivatives, obtained with differential scanning calorimetry (DSC, see Methods and Supporting Information S9 for details), are shown in Table \ref{tbl:Tg}. While  \textbf{1} exhibits a reasonable $T_g$ \cite{Li_Duan_Liang_Han_Wang_Ye_Liu_Yi_Wang_2017}($\sim$110°C), the highest performing derivative, \textbf{4}, has a lower $T_g$ (76°C), potentially limiting its long-term operational stability and integration with high-temperature fabrication steps.  To directly probe the thermal stability at high temperature, we measured the nonlinear response of compounds \textbf{4 }and \textbf{5} during heating. The films exhibited the remarkable stability, maintaining their nonlinear response to temperatures exceeding their $T_g$, a characteristic typical of ultrastable glasses\cite{Dalal_Walters_Lyubimov_dePablo_Ediger_2015,Swallen_Kearns_Mapes_Kim_McMahon_Ediger_Wu_Yu_Satija_2007}, before the orientational order was irreversibly lost at a higher temperature (see Supporting Information S10, Figure S21). While these results demonstrate excellent short-term thermal robustness, and the films showed no signs of degradation after several months at room temperature, the long-term operational stability  near or above the glass transition temperatures remains an important area for future study. Future molecular design efforts should strategically co-optimize both the nonlinear optical performance and thermal stability to ensure robust and practical device implementation.

\begin{table}[H]
  \caption{Glass transition temperature of the compounds. The glass transition temperature $T_g$ of compound \textbf{1} was reported in Ref. \citenum{Li_Duan_Liang_Han_Wang_Ye_Liu_Yi_Wang_2017}. Other $T_g$ values were determined by DSC.}
  \label{tbl:Tg}
  \begin{tabular}{cccc}
    \hline
Compound  & $T_g$ [$^\circ$C] \\
    \hline
   \textbf{1}    &110\\
    \textbf{2}&103\\   
    \textbf{3}&102\\
    \textbf{4}&76 \\
    \textbf{5}&90\\
    \hline
  \end{tabular}
\end{table}

\section{Conclusion}
In summary, we have designed, synthesized, and characterized a series of TPA-QCN derivatives to enhance the second-order nonlinearity of spontaneously oriented organic thin films prepared by vacuum evaporation. By employing strategies targeting both molecular hyperpolarizability and molecular orientation, we achieved significant improvements over the parent compound. Notably, intramolecular bridge-locking in compound \textbf{4} resulted in a threefold enhancement near resonance ($\lambda_{pump}=1266$~nm) and a twofold enhancement off-resonance ($\lambda_{pump}=1550$~nm), yielding  $\chi^{(2)}_{31}=41.6\pm0.3$ pm V$^{-1}$ and $\chi^{(2)}_{33}=44\pm3$ pm V$^{-1}$  at 1266 nm, competitive with established inorganic materials like but achieved via a simple, scalable, non-epitaxial and poling-free fabrication process. Our analysis, combining nonlinear optical measurements with computed hyperpolarizabilities, GSP, and optical anisotropy, strongly indicates that optimizing molecular orientation, rather than solely maximizing hyperpolarizability, was the key factor driving the observed performance enhancements, particularly for compounds \textbf{4} and \textbf{5}.

These findings underscore the potential of molecular engineering to control and enhance $\chi^{(2)}$ properties in spontaneously oriented films, positioning them as a promising platform for nonlinear optical applications where the increased fabrication complexity, footprint and optical losses associated with poling electrodes are a critical limitation. However, they also highlight the need for a deeper fundamental understanding of the complex interplay between molecular structure, intermolecular forces, and deposition kinetics that govern spontaneous orientation during PVD. Future progress towards even higher nonlinearities and practicality will likely require strategies that simultaneously optimize  $\beta$, orientation and thermal stability. Furthermore, exploring the impact of deposition parameters (\textit{e.g.}, rate, substrate temperature)  and film composition (\textit{e.g.}, dilution in host matrices)\cite{Pakhomenko_He_Holmes_2023,Hofmann_Schmid_Brütting_2021,Tanaka_Auffray_Nakanotani_Adachi_2022,hofmann2025enhancement, Noguchi_Osada_Ninomiya_Gunawardana_Koswattage_Ishii_2021,Gunawardana_Osada_Koswattage_Noguchi_2021}, which are known to influence molecular orientation but were beyond the scope of this initial study, represents a promising avenue for further optimization.

\section{Methods}
\subsection{Synthesis}

\begin{figure}[H]
    \centering
    \includegraphics[width=0.6\linewidth]{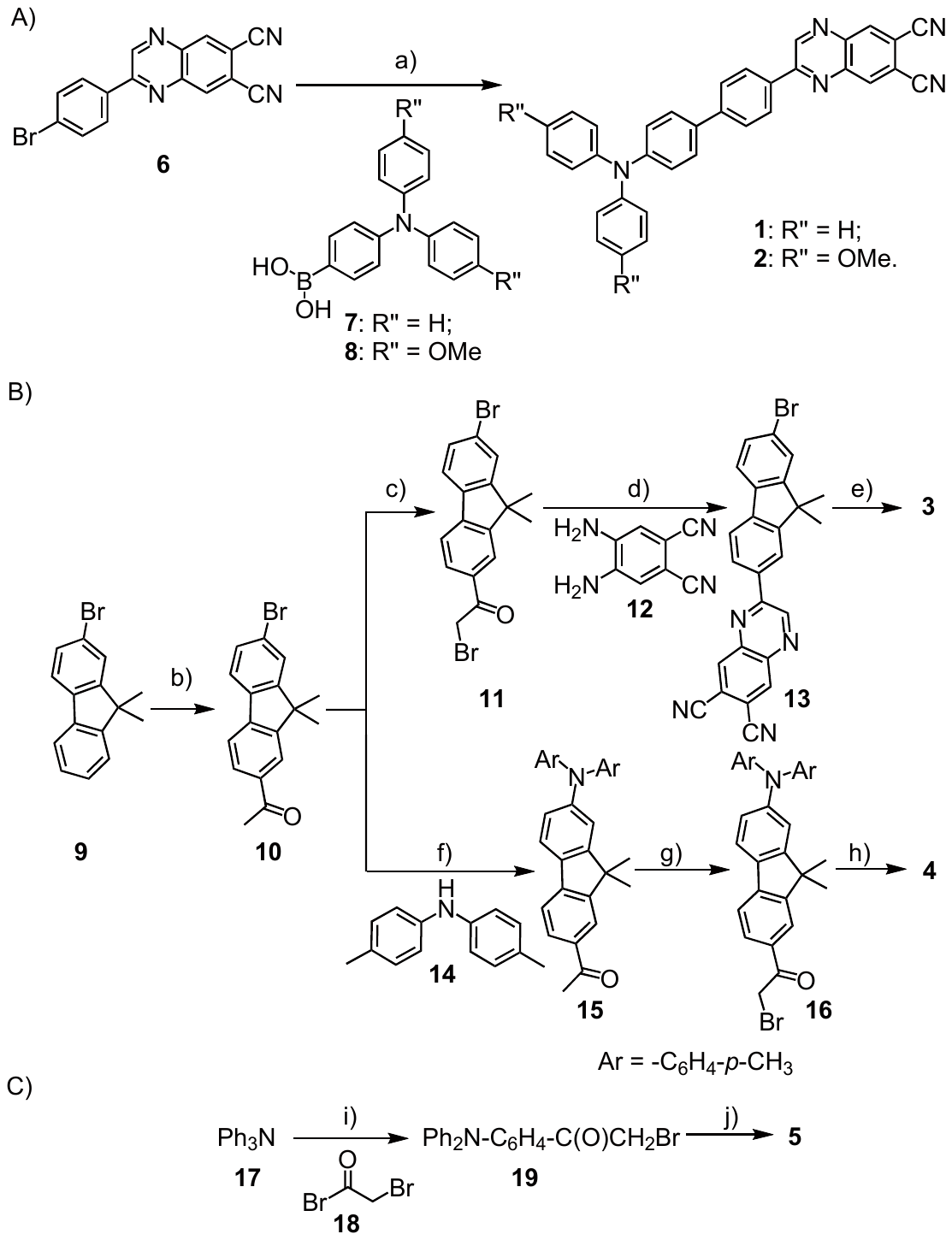}
    \caption{Synthetic schemes for the synthesis of compounds \textbf{1}--\textbf{5}.
Reagents and conditions: 
(a) \textbf{7}, K$_2$CO$_3$, Pd(PPh$_3$)$_4$, H$_2$O, toluene, reflux, (29\%). 
(b) AcCl, AlCl$_3$, CH$_2$Cl$_2$, rt, (91\%). 
(c) NBS, TsOH, MeCN, reflux, (91\%). 
(d) \textbf{12}, CTAB, H$_2$O, reflux, (88\%). 
(e) \textbf{7}, K$_2$CO$_3$, Pd(PPh$_3$)$_4$, H$_2$O,/toluene, THF,  reflux, (74\%). 
(f) \textbf{14}, Cs$_2$CO$_3$, Pd(OAc)$_2$, P($t$Bu)$_3$, toluene, reflux, (74\%).  
(g) CuBr$_2$, EtOAc, reflux, (63\%). 
(h) \textbf{12}, CTAB, H$_2$O, reflux, (45\%). 
(i) \textbf{18}, AlCl$_3$, CH$_2$Cl$_2$, rt, (71\%). 
(j) \textbf{12}, CTAB, H$_2$O, reflux, (56\%). }
    \label{Fig1:Synthesis}
\end{figure}

Synthetic schemes are shown on Figure \ref{Fig1:Synthesis}. Synthesis of \textbf{1} was reported before\cite{Li_Duan_Liang_Han_Wang_Ye_Liu_Yi_Wang_2017}. Its dimethoxy derivative \textbf{2} was synthesized by Suzuki cross-coupling of boronic acid \textbf{8} with brominated quinoxaline synthon \textbf{6} (Figure \ref{Fig1:Synthesis}A).\cite{Li_Duan_Liang_Han_Wang_Ye_Liu_Yi_Wang_2017} Fluorene-bridged chromophores \textbf{3} and \textbf{4} were obtained in four steps starting with acetylation of 2-bromo-9,9-dimethylfluorene \textbf{9} (Figure \ref{Fig1:Synthesis}B). The resulting \textbf{10} was brominated with NBS followed by cyclization with 4,5-diaminophthalonitrile forming arylbromide \textbf{13} which underwent Suzuki coupling with triphenylamineboronic acid \textbf{7} to afford chromophore \textbf{3}. Synthesis of \textbf{4} required a different synthetic procedure starting from the diarylamino donor synthon rather than QCN because the attempts of Buchwald-Hartwig amination of \textbf{13} have failed. Amination of bromofluorene \textbf{10} with di(\textit{p}-tolyl)amine \textbf{14} resulted in the intermediate \textbf{15} which was brominated with $\text{CuBr}_2$ followed by the mentioned above cyclization with \textbf{12} to afford chromophore \textbf{4}. Synthesis of smallest chromophore \textbf{5} (Figure \ref{Fig1:Synthesis}C) was accomplished by acylation of triphenylamine \textbf{17} with bromoacetylbromide \textbf{18} followed by cyclization with\textbf{ 12}. The detailed synthetic procedures and the characterization of all compounds are described in the Supporting Information (S11, S12 and S14).

\subsection{Quantum Chemical Calculations}
Density functional theory (DFT) calculations were performed using the Gaussian 16 software package. Geometries were optimized using the M062X hybrid functional and 6-31G(d) basis set, which was selected based on a benchmark against other functionals (see Supporting Information, Section S2). The static ($\beta_0$) and frequency-dependent first hyperpolarizabilities were calculated at the same level of theory.Time-dependent DFT (TDDFT) was used to calculate the excitation energies and oscillator strengths of transitions.

\subsection{Thin Films Fabrication}
Organic thin films were prepared using thermal evaporation in a high-vacuum chamber with a base pressure maintained below $1 \times 10^{-6}$~mbar. Fused silica and silicon wafers coated with a 100~nm thermal oxide layer served as substrates. Prior to deposition, substrates were sequentially cleaned by sonication in detergent solution, deionized water, acetone, and isopropanol, followed by immediate drying under a nitrogen stream and treatment with UV-Ozone (15 minutes) to ensure surface cleanliness and hydrophilicity. The purified TPA-QCN derivative materials (\textbf{1}--\textbf{5}) were loaded into resistively heated crucibles. The deposition rate was monitored in situ using a quartz crystal microbalance and maintained within the range of 0.5--2.0~Å/s. During evaporation, the substrates were kept at ambient temperature without active heating or cooling. Final film thicknesses, ranging from 49~nm to 108~nm, were determined ex situ using spectroscopic ellipsometry.
\subsection{Linear Optical characterization}
Absorbance spectra of the films deposited on fused silica were acquired using a dual-beam ultraviolet-visible-near infrared (UV-Vis-NIR) Perkin-Elmer Lambda 25 spectrophotometer over a spectral range of 200~nm to 1100~nm. Variable Angle Spectroscopic Ellipsometry (VASE) measurements were performed using a rotating-compensator Woolam RC2 ellipsometer  to determine the optical constants and thicknesses of the films. Muller matrix parameters were collected over a wavelength range from 210~nm to 2500~nm at multiple angles of incidence (e.g., 45 $^\circ$,55$^\circ$, 65$^\circ$ and 75$^\circ$). The acquired data were analyzed  to fit the experimental spectra and extract $n(\lambda)$ and $k(\lambda)$. The models accounted for uniaxial anisotropy, assuming the optic axis was perpendicular to the substrate plane. Detailed optical constants spectra and details about the modelling are provided in the Supporting Information S5.
\subsection{Second-Harmonic Polarimetry}
The second-order nonlinear optical response was characterized using second-harmonic polarimetry. The fundamental pump beam was provided by a femtosecond laser (Light Conversion Pharos) system coupled to an optical parametric amplifier (OPA, Light Conversion Orpheus), delivering pulses at wavelengths ($\lambda_{\text{pump}}$) of 1266~nm and 1550~nm with a repetition rate of  100~kHz and pulse duration of approximately 170 fs (characterized using frequency-resolved optical gating\cite{kane1993characterization}). A half-wave plate mounted on a motorized rotation stage was used to precisely control the polarization state of the incident beam. The beam was focused onto the sample using a lens (f=200~mm), resulting in a beam waist of approximately 30~$\mu$m at the sample plane (measured using a camera). A long-pass filter positioned before the sample blocked any pre-generated SHG signal from upstream optics. Samples were mounted on a rotation stage to allow measurements at multiple angles of incidence (typically 45$^\circ$ to 70$^\circ$, with data taken at a minimum of two angles per sample). The generated second-harmonic signal transmitted through the sample was collimated, spectrally filtered using a combination of short-pass filters to reject the fundamental pump wavelength, and passed through a polarization analyzer set to select either the TE or TM component. The intensity of the selected second-harmonic component was measured using an intensity-calibrated  CMOS camera. For each incidence angle, the SHG intensity was recorded as a function of the incident polarization angle (controlled by the rotating half-wave plate) for both TE and TM output polarizations. System calibration and validation were performed using a standard z-cut $\alpha$-quartz reference sample.

\subsection{Nonlinear Coefficients Extraction}
The second-order nonlinear susceptibility tensor elements ($\chi^{(2)}_{31}$ and $\chi^{(2)}_{33}$, assuming $C_{\infty v}$ symmetry) were extracted from the angle-dependent SHG polarimetry data using nonlinear transfer matrix calcultations \cite{Bethune_1991} . This model calculates the expected second-harmonic field for a given incident polarization, angle of incidence, and output polarization. The model incorporates the experimentally determined film thickness and complex refractive indices ($n(\lambda)$, $k(\lambda)$) at both the fundamental ($\omega$) and second-harmonic ($2\omega$) frequencies, fully accounting for linear absorption, dispersion, and multiple reflection/interference effects within the layers. The undepleted pump approximation was employed, justified by the low ($<$1\%) SHG conversion efficiencies observed. A least-squares fitting algorithm was used to simultaneously fit the model predictions to the experimental TE- and TM-polarized SHG intensity patterns (as shown in Figure~\ref{Fig2}b,c), varying the values of $\chi^{(2)}_{31}$ and $\chi^{(2)}_{33}$ to minimize the difference between calculated and measured signals. The final reported $\chi^{(2)}$ values represent the uncertainty-weighted means obtained from fits to data acquired at multiple incidence angles for each sample.

\subsection{Surface Potential Measurement}
Surface potential measurements were performed using a Trek 320C electrostatic voltmeter under ambient conditions in the dark immediately following film deposition to minimize the effects of ambient light and potential photocarrier generation. Measurements were conducted on samples deposited on silicon substrates with 100~nm thermal oxide, where approximately half of the substrate was physically masked during evaporation to provide an uncoated reference area adjacent to the organic film. The surface potential difference  between the organic film surface and the uncoated reference region was measured. Data were acquired from 2 to 4 independent samples for each compound to ensure reproducibility. The net surface potential was divided by the film thickness (determined by ellipsometry) to obtain the Giant Surface Potential slope.

\subsection{Thermal characterization}
Thermogravimetric analyis (TGA) measurements were carried out on a TGA-Q500 analyzer under N2 atmosphere at a heating rate of 20 °C min$^{-1}$ unless otherwise noted. DSC measurements were carried out on a DSC-2500 analyzer with Tzero aluminum pin hole hermetic pan under $\text{N}_2$. To assess the effects of previous thermal history on subsequent thermal events, heat-cool cycle method (-20 °C to 260 °C) at a rate of 10 °C $^{-1}$ was used.

\begin{acknowledgement}
P.L.T. acknowledges support from the Trottier Energy Institute and Natural Science and Engineering
Research Council of Canada Graduate Scholarship. H.V.H. and Z.H acknowledge support from the Fonds de Recherche du Québec - Nature et Technologies Postdoctoral Fellowship. S.K.C. acknowledges support from the Canada Research Chair Program. This work was funded by the
Natural Science and Engineering Research Council Quantum
Consortium Program (Quantamole).The authors acknowledge support from Digital Research Alliance of Canada (www.alliancecan.ca) for enabling the DFT simulations. 
\end{acknowledgement}

\begin{suppinfo}
The following sections are available in the Supplementary Information: 
Discussion on molecular structure approximations, computational method details and additional results, measurements of the films' density, XRD characterization, refractive index data and ellipsometry modeling details, near-infrared transmission spectra,  two-level enhancement factor, calculations of $\langle cos\theta \rangle$, thermal characterization data, detailed synthesis procedures, compound characterization data.

\end{suppinfo}

\bibliography{Main}
\end{document}